\documentclass{ws-procs9x6}            
\begin{document}
\title{Low Energy Magnetic Radiation (LEMAR ) of Warm Nuclei}

\author{\underline{S.Frauendorf$^*$} }

\address{University Notre Dame,\\
Notre Dame, IN 46556, USA\\
$^*$E-mail: sfrauend@nd.edu}
\author{B. A. Brown}
\address{ National Superconducting Cyclotron Laboratory and Department of
             Physics and Astronomy, Michigan State University,\\
             East Lansing,  Michigan 48824, USA}
\author{R. Schwengner}
\address{ Institut f\"ur Strahlenphysik, Helmholtz Zentrum Dresden Rossendorf,\\
 01328 Dresden, Germany}

\begin{abstract}
The enhancement observed below 2 MeV  in the radiative  strength function of nuclei near closed shells is 
 explained by shell model calculations as M1  transitions between excited states. In the open-shell a
 change to a bimodal structure composed of the zero energy spike and a scissors resonance is found.
 The features are caused by realignment of high-j orbitals.

\end{abstract}

\keywords{radiative strength, shell model, high-j orbitals.}

\bodymatter
\bigskip
\bigskip
\section{Introduction}
 Photonuclear reactions and the inverse radiative-capture reactions between
nuclear states in the region of high excitation energy and large level density
are of considerable interest in many
applications. 
Radiative neutron capture, for example, plays a central role in
the synthesis of the elements in various stellar environments, for next-generation nuclear technologies, and
as the transmutation of long-lived nuclear waste. 

\begin{figure}[t]
 \includegraphics[width=\linewidth]{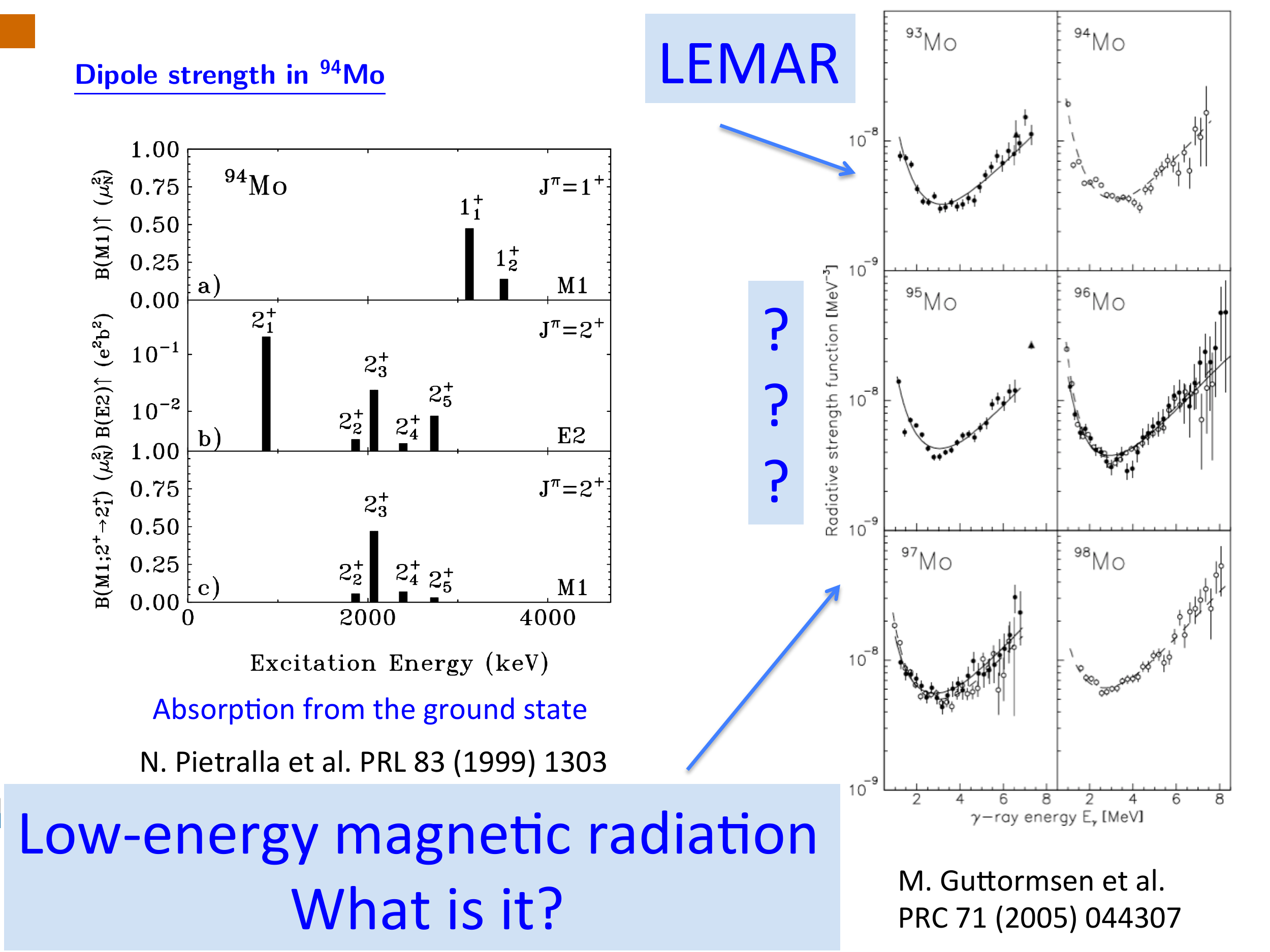}
 \caption{ \label{f:gsAbsorption} (Color online) 
Left: Reduced absorption probabilities from the ground state by $(\gamma,\gamma')$ experiments \cite{piet99}. Right:
Strength function for $\gamma$ de-excitation after the  ($^3$He,$^3$He') reaction \cite{gut05}.}
\end{figure}

A critical input to
calculations of the reaction rates is the average strength of the cascade of 
$\gamma$-transitions  de-exciting the nucleus,
which is described by the radiative  strength function.
 An increase of the
dipole strength function below 3 MeV toward low $\gamma$-ray energy has recently been
observed  in  nuclides in the mass range from $A \approx$ 40 to 100. As an example,  Fig. \ref{f:gsAbsorption} 
shows  the $\gamma$ strength function obtained from
($^3$He,$^3$He') reactions on various Mo isotopes. The enhancement is not observed in the inverse process of absorbing $\gamma$-quanta
by nuclei in the ground state. 
Only few discrete lines are found within the interval of the first 4 MeV. The question arises which is the origin of the radiation enhancement and how does it
depend on the $N$ and $Z$. Refs. \cite{lar10,sim16} showed that the low-energy enhancement may increase the neutron capture rate in the stellar r-process environment 
up to a factor 100-1000.  The dominant dipole character of the low-energy strength
was demonstrated in Ref.~\cite{lar13L} and an indication for an M1 character
 was discussed for the case of $^{60}$Ni \cite{voi10}.

\section{The LEMAR spike}
\begin{figure}[t]
\begin{center}{\includegraphics[width=0.48\linewidth]{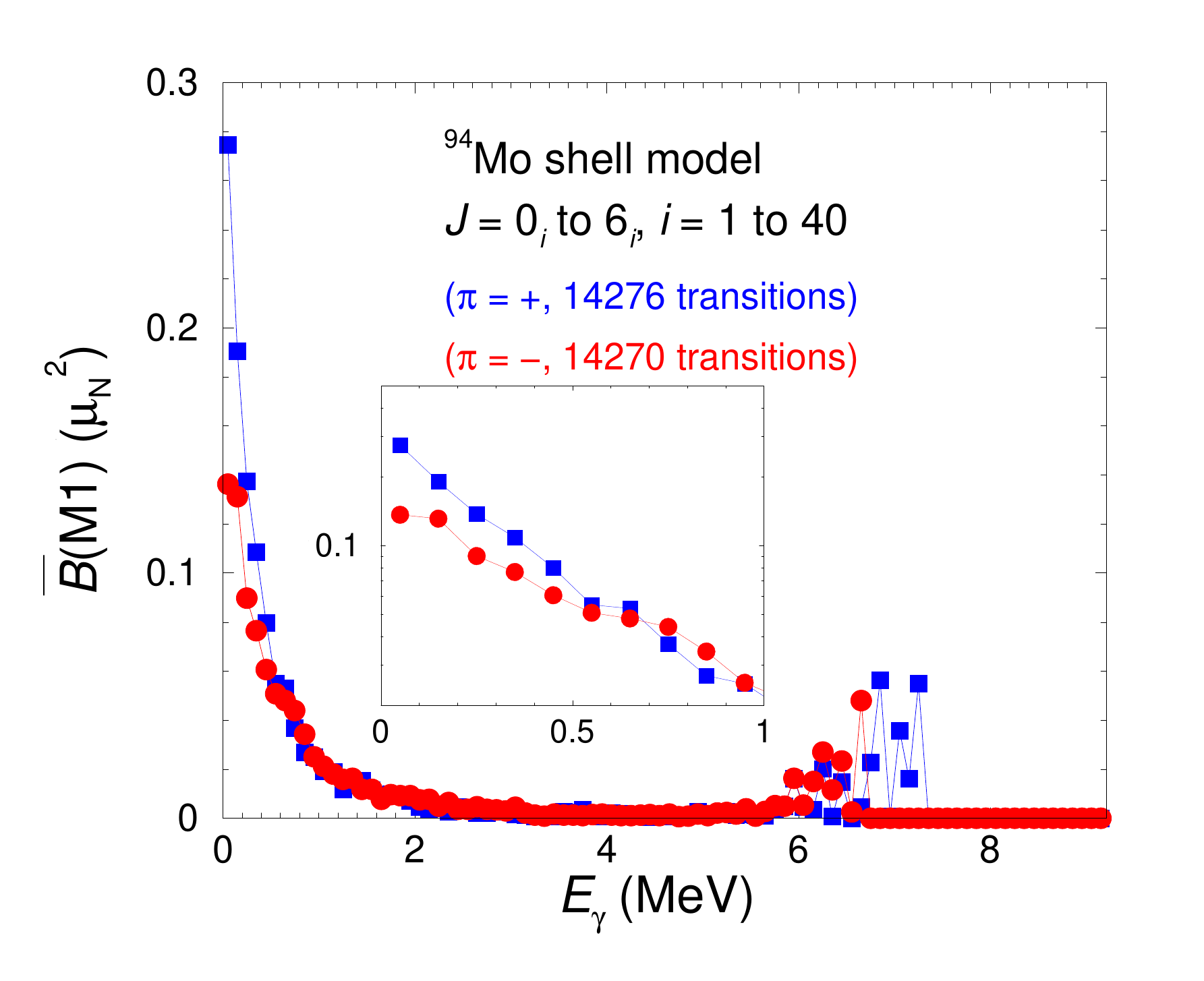}}
\includegraphics[width=0.51\linewidth]{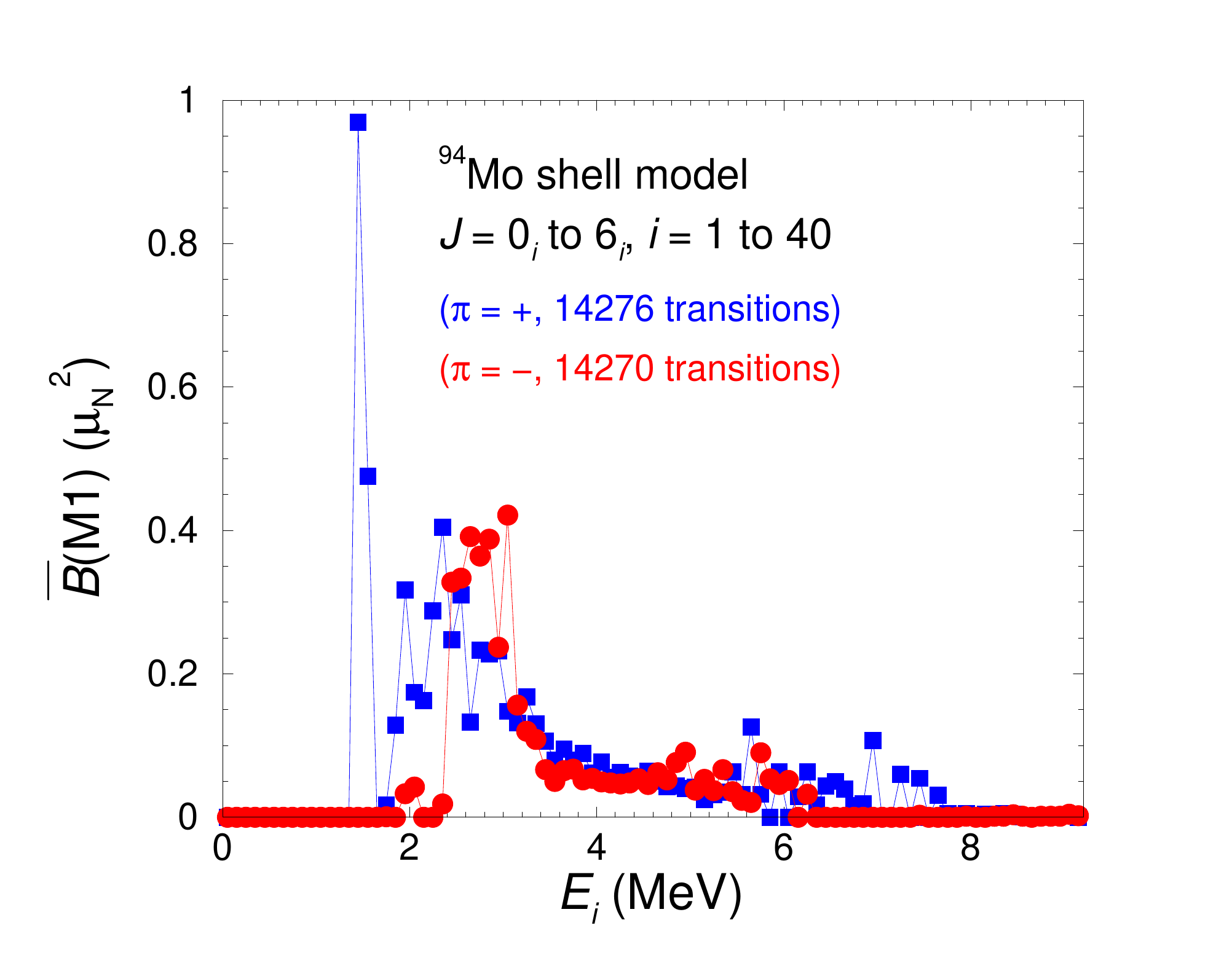}
\end{center}
 \caption{\label{fig:94Mo}(Color online) Left: Average $B(M1)$ values in 100 keV
bins of transition energy calculated for positive-parity (blue squares) and
negative-parity (red circles) states in $^{94}$Mo. The inset shows the
low-energy part in logarithmic scale. Right: Average $B(M1)$ values in 100 keV
bins of excitation energy calculated for positive-parity (blue squares) and
negative-parity (red circles) states in $^{94}$Mo. From \cite{sch13}.
}
\end{figure}

Our Shell Model studies of the Mo isotopes explained the enhancement as Low Energy MAgnetic Radiation (LEMAR) generated by transitions between 
excited states few MeV above yrast \cite{sch13}. The Shell Model calculations  included the active proton orbits
$\pi(0f_{5/2}, 1p_{3/2}, 1p_{1/2}, 0g_{9/2})$ and the neutron orbits
 $\nu(0g_{9/2}, 1d_{5/2}, 0g_{7/2})$ relative to a $^{68}$Ni core. A set of empirical matrix elements  for the effective
interaction and   effective $g$-factors of $g^{\rm eff}_s = 0.7  g^{\rm free}_s$ have been used. 
The calculations included states with spins from $J$ = 0 to 6 for $^{90}$Zr and
$^{94}$Mo and from $J$ = 1/2 to 13/2 for $^{95}$Mo. For each spin the lowest 40
states were calculated. The reduced transition probabilities $B(M1)$  were
calculated for all transitions from initial to final states with energies
$E_f < E_i$ and spins $J_f = J_i, J_i \pm 1$. This resulted in more than 14000 $M1$ transitions for each parity
$\pi = +$ and $\pi = -$, which were sorted into 100 keV bins according to
their transition energy $E_\gamma = E_i - E_f$. The average $\overline{B}(M1,E_\gamma)$ value for
one energy bin was obtained as the sum of all $B(M1)$ values divided by the
number of transitions within this bin. The results for 
$^{94}$Mo are shown in Fig.~\ref{fig:94Mo}. 
The left panel demonstrates appearance of the LEMAR spike, which exponentially decreases with the transition energy $E_\gamma$.
The  exponential dependence on the transition energy is retained by
the M1 strength functions, which are defined by the relation 
\begin{equation}
f_{M1}(E_\gamma)
= 16\pi/9 (\hbar c)^{-3}\overline{B}(M1,E_\gamma)\rho(E_i),
\end{equation}
where the level density at the initial state  $\rho(E_i)$ is obtained from the Shell Model calculations.

\begin{figure}[t]
\begin{center}{\includegraphics[width=\linewidth]{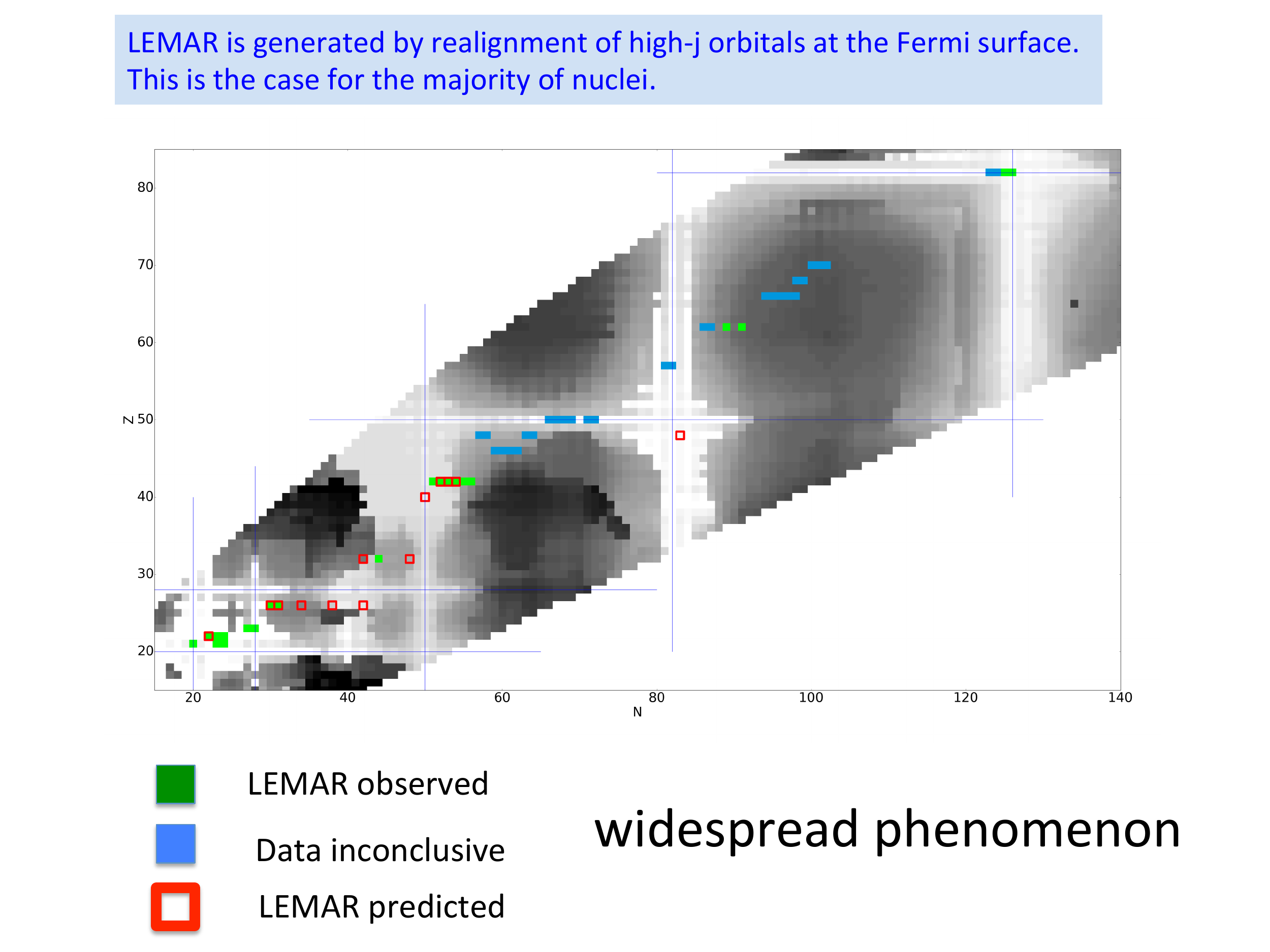}}
\end{center}
 \caption{\label{f:Landscape}(Color online) Experimental presence of the LEMAR spike and predictions by 
 Shell Model calculations. }
 \end{figure}

As seen in the right panel of Fig.~\ref{fig:94Mo}, LEMAR is generated by transition originating from states between 2 and 6 MeV.  
Inspecting the composition of initial and final states, we found large $B(M1)$ values for transitions
between states that contain a large component (up to about 50\%) of the same
configuration with broken pairs of both protons and neutrons in high-j
orbits. The largest $M1$ matrix elements connect configurations with the spins
of $g_{9/2}$ protons re-coupled with respect to $g_{9/2}$ neutrons to
the total spin $J_f = J_i, J_i \pm 1$. These orbitals have large magnetic moments.
Large M1 matrix elements connect configurations which differ by only the
angular momentum projection of these high-j orbitals. Our Shell Model studies of $^{56,57}$Fe  \cite{bro14Fe} 
and $^{60,64,68}$Fe \cite{sch17}
 found also the LEMAR spike, which is generated by reorientation of
$f_{7/2}$ proton holes and $g_{9/2}$ neutrons.  High-j particle or hole orbitals with $j=l+1/2$ have the 
largest magnetic moments. As they 
appear near the Fermi level all over the nuclear chart one expects that LEMAR spike is omnipresent as well.

\begin{figure}[t]
\begin{center}
\includegraphics[width=\linewidth]{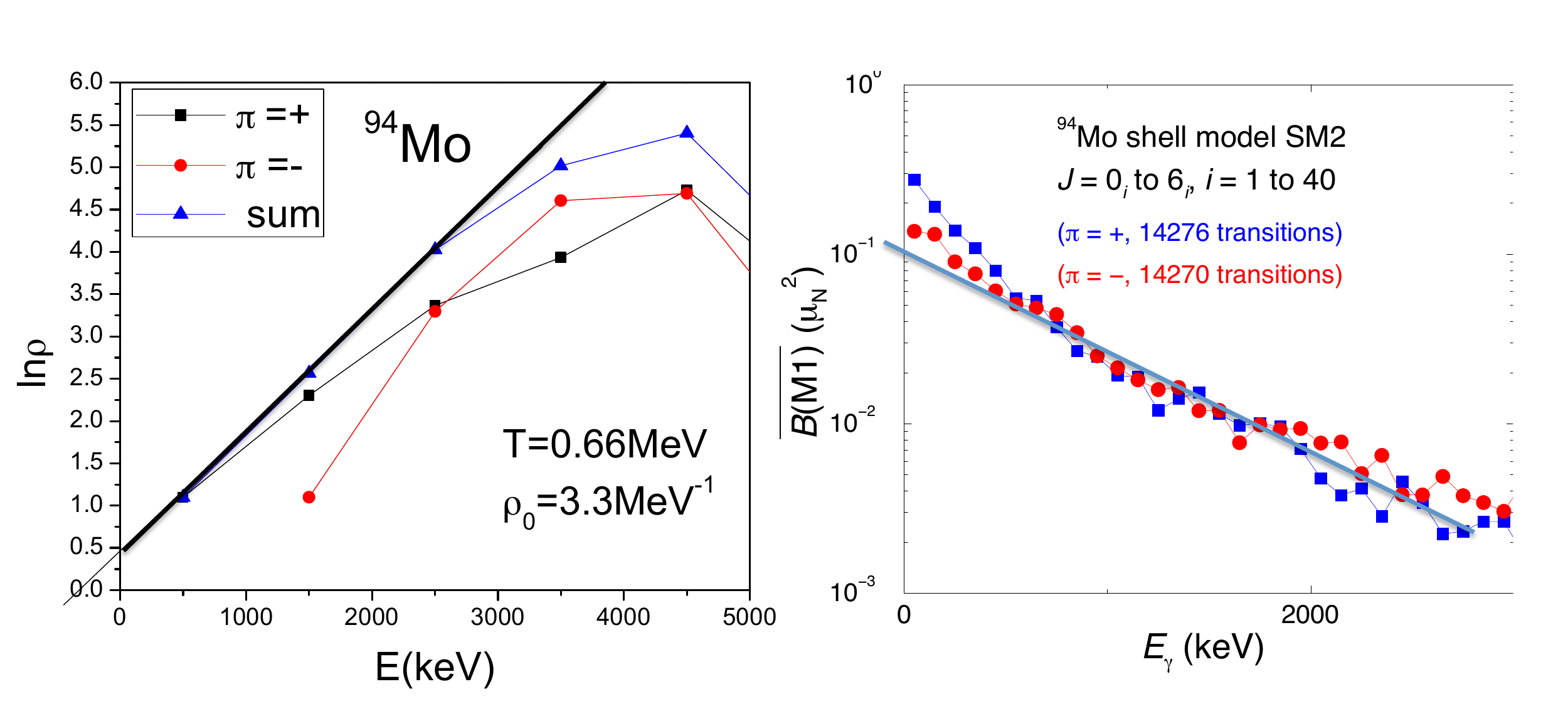}
\end{center}
 \caption{\label{f:TrhoTM1} (Color online) Left: Shell Model calculation \cite{sch13} of the level density (left) and average $B(M1)$ values of
$^{94}$Mo.   The straight lines have the  slope $\pm 1/T$, $T=0.66$ MeV. }
 \end{figure}

The left panel of Fig. \ref{f:TrhoTM1} shows the logarithm of the level density. At low energy the level density follows the constant temperature
expression $\rho(E)=\rho_0\exp(E/T)$.  Above 4 MeV the level density is underestimated because of the  truncation of the configuration space.   
The right panel demonstrates that the exponential decrease of the LEMAR spike is consistent with $T$ derived from the level density. For the Mo isotopes
 there is  consistency between the value of $T\approx 0.8$ MeV derived from the experimental level densities \cite{chang06} and from the LEMAR
 spike of the $\gamma$ strength functions \cite{gut05}. As illustrated by Fig. \ref{f:Sm153} below,  the consistency 
 also holds for the experimental level densities and $\gamma$
 strength functions of $^{151,153}$Sm.

\section{Consequences of deformation and properties of the Scissors Resonance}
  In our Shell Model studies of  $^{60,64,68}$Fe \cite{sch17} we investigated the influence of quadrupole correlation on LEMAR.
  The calculations were carried out in the CA48PN model space with the CA48MH1
Hamiltonian using the code NuShellX@MSU. The model
space included the 
$\pi(0f_{7/2}^{(6-t)}, 0f_{5/2}^t, 1p_{3/2}^t, 1p_{1/2}^t)$ proton orbits with
$t$ = 0, 1, 2, and the 
$\nu(0f_{5/2}^{n5}, 1p_{3/2}^{p3}, 1p_{1/2}^{p1}, 0g_{9/2}^{g9})$ neutron orbits.
The calculations of M1 strengths were carried out in the same way as for the Mo isotopes. 
They included the lowest 40 states each with
spins from $J_i, J_f$ = 0 to 10, which resulted in more than 22000 M1 transitions for each parity.
 Effective $g$ factors of $g^{\rm eff}_s = 0.9  g^{\rm free}_s$ were applied. 

 \begin{figure}[t]
\begin{center}{\includegraphics[width=\linewidth]{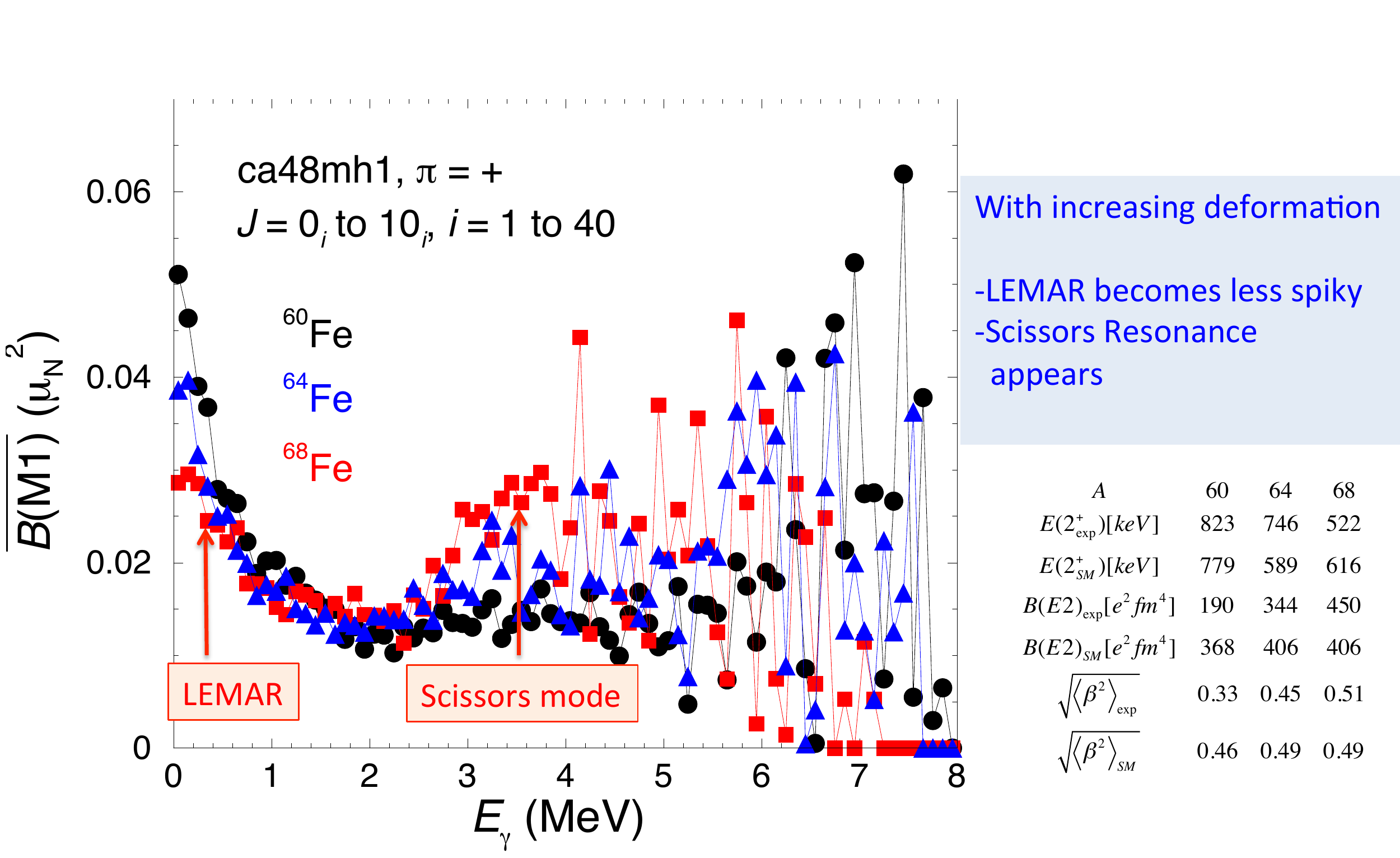}}
\end{center}
 \caption{\label{f:BM1Fe} (Color online) Calculated average B(M1) values for the positive parity states in $^{60,64,68}$Fe. The table compares the
calculated  properties of the first $2^+$ state with experiment.  }
 \end{figure}
 
  The comparison with the experimental $B(E2,2^+_1\rightarrow 0^+_1)$ 
values in the Table included in  Fig. \ref{f:BM1Fe} shows that the calculations reproduce the 
quadrupole collectivity in the considered isotopic chain, in particular the
enhancement observed around $N$ = 40.
The respective experimental ratios $E(4^+)/E(2^+)$=2.57,  2.36,  2.66, which   
are well accounted for by the calculations $E(4^+)/E(2^+)$=2.56, 2.68, 2.43,  
indicate the transitional character of the ground state bands of all considered isotopes. 

The comparison of the
$\overline{B}(M1)(E_\gamma)$ values of the various isotopes shows that the
shape of the distributions changes when going from $N$ = 34, four neutrons
above the closed shell, to $N$ = 42, the middle of the $fpg$ shell. One
observes a weakening of the LEMAR spike and the development of a bump in the
range from about 2 to 5 MeV, which is most pronounced in $^{68}$Fe. We
interpret this bump as the Scissors Resonance (SR) built on excited states.

 \begin{figure}[t]
\begin{center}{\includegraphics[width=\linewidth]{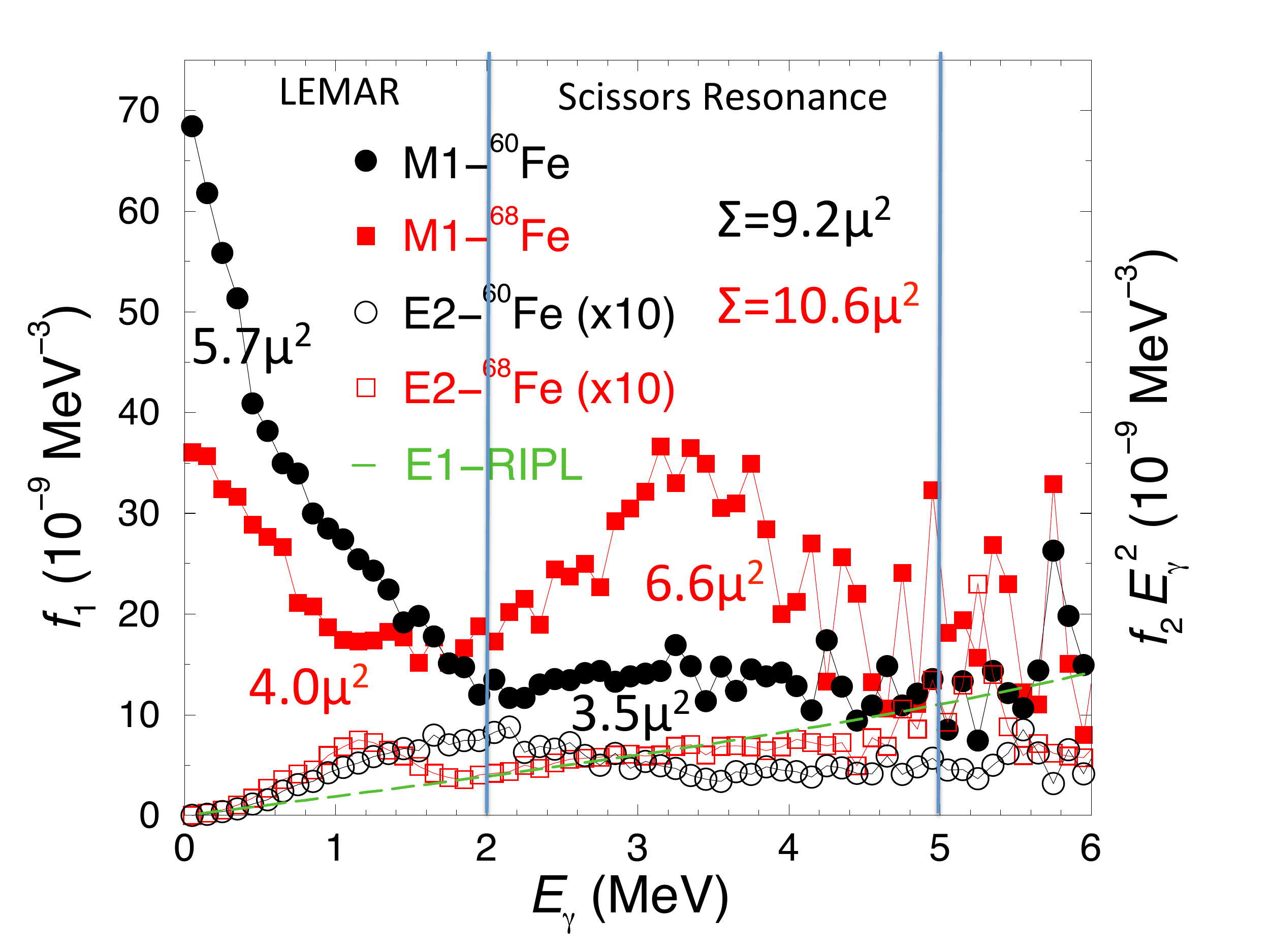}}
\end{center}
 \caption{\label{f:f1Fe} (Color online) Calculated M1 strength functions ($f_1$)  and E2 strength functions $\left(f_2E_\gamma^2\right)$  in 
 $^{60,68}$Fe. Note the factor of 10 for  E2. The numbers quote the integrated strength for the LEMAR spike
$\left(\int_0^2 f_1(E)dE\right)$ 
and the Scissors Resonance $\left(\int_2^5 f_1(E)dE\right)$. The
 vertical lines indicate the respective integration regions. For $^{68}$Fe the sum of the M1 strength from the ground state to all 1$^+$ states is 
 $\sum B(M1,0^+_1\rightarrow 1^+)=1.7\mu^2$.  }
 \end{figure}
 
  \begin{figure}[t]
\begin{center}
\includegraphics[width=\linewidth]{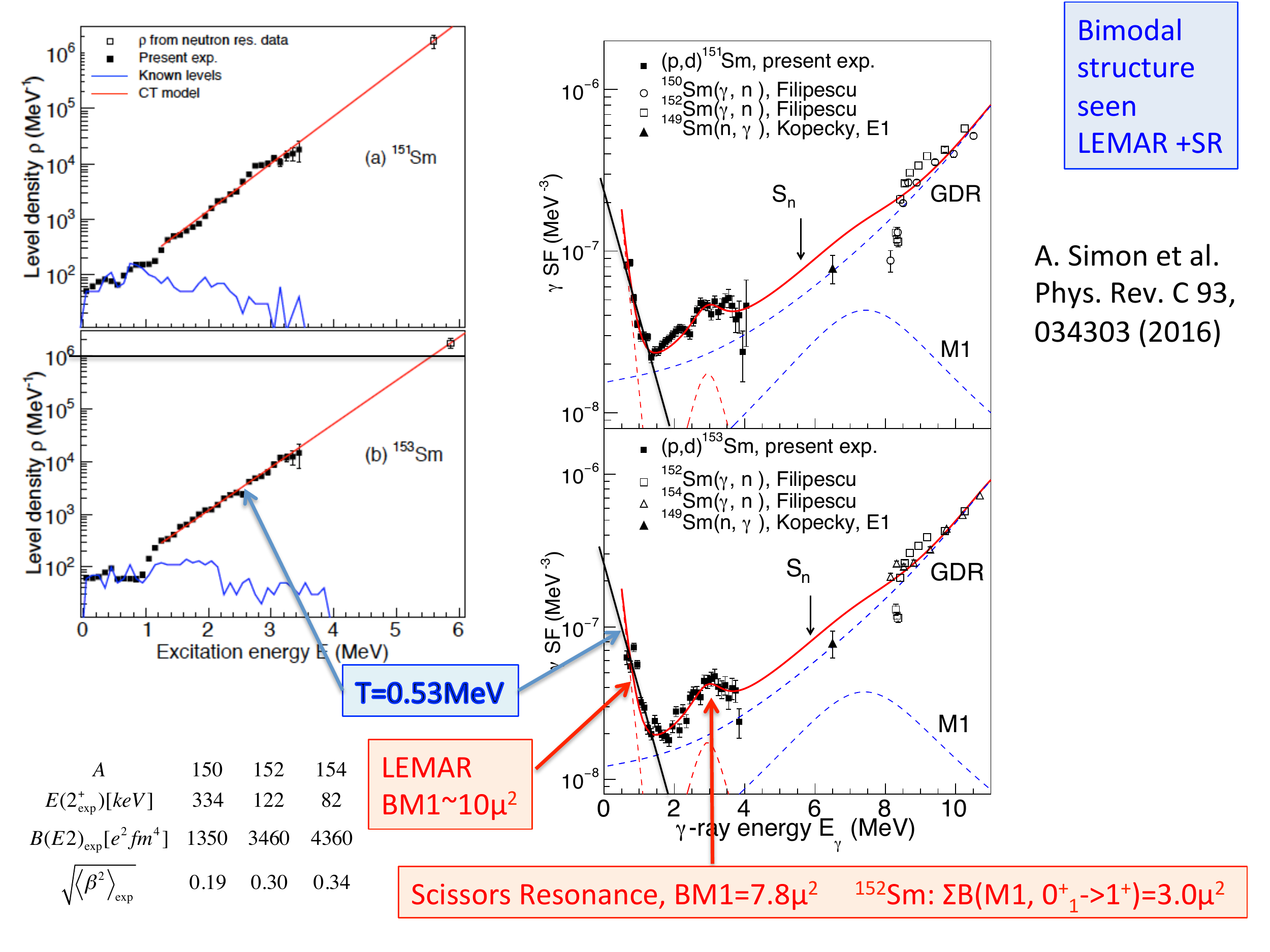}
\end{center}
 \caption{\label{f:Sm153} (Color online) Experimental level densities (left) and $\gamma$ strength functions (right) of
 $^{151,153}$Sm measured and derived by means of the Oslo method \cite{sim16}. 
 The dashed lines show a fit of three Lorentzians (GDR, M1GR, SR) to the $\gamma$ strength functions.
 The temperature $T$ derived from the slope of the level density (left) is used to extrapolate the LEMAR spike to $E_\gamma=0$.
The sum of the M1 strength from the ground state to all 1$^+$ states is quoted at the bottom.  }
 \end{figure}

 Fig. \ref{f:f1Fe} shows the M1 strength functions for $^{60,68}$Fe, which were calculated using level densities from the Shell Model  calculations.
 Like the average $\overline B(M1)$ values they develop the bimodal LEMAR-SR structure 
 when going to the middle of the open shell. The integrated  strength deduced from the
strength functions 
below 5 MeV varies by only 8\% at most from an average of 9.80 $\mu^2_N$.
The slight increase with $N$ is attributed to the progressive occupation of the
$g_{9/2}$ shell for $N$ = 34, 38, 42. That is, the bimodal structure develops with increasing $N$ 
by shifting  strength  from  the LEMAR spike to the SR while the sum stays nearly constant. 

For comparison, the standard E1 strength function with parameters taken from 
the RIPL data base  and E2 strength functions $f_2$ obtained from
the present calculations  are also shown in
Fig.~\ref{f:f1Fe}. The latter were multiplied with $E_\gamma^2$ to be
directly comparable with the dipole strength functions $f_1$. The E2 strength
$f_2 E_\gamma^2$ is more than one order of magnitude smaller and the 
E1 strength exceeds the M1 strength only at energies greater than 6 MeV.

The strength functions deduced  by means of the Oslo method from  high-resolution
experiments  by A. Simon  {\it et al.}  \cite{sim16} showed for the first time the LEMAR spike 
down to about $E_\gamma$ = 1 MeV as well as a SR around 3 MeV in one
nuclide. Fig. \ref{f:Sm153} displays the results together with a table of the properties  of the $2^+$ states of the neighbors, which 
demonstrate the well deformed nature of the nuclides. Thus we conclude that the bimodal LEMAR-SR structure is a general feature 
expected for all  deformed nuclei.

 \begin{figure}[t]
\begin{center}
\includegraphics[width=\linewidth]{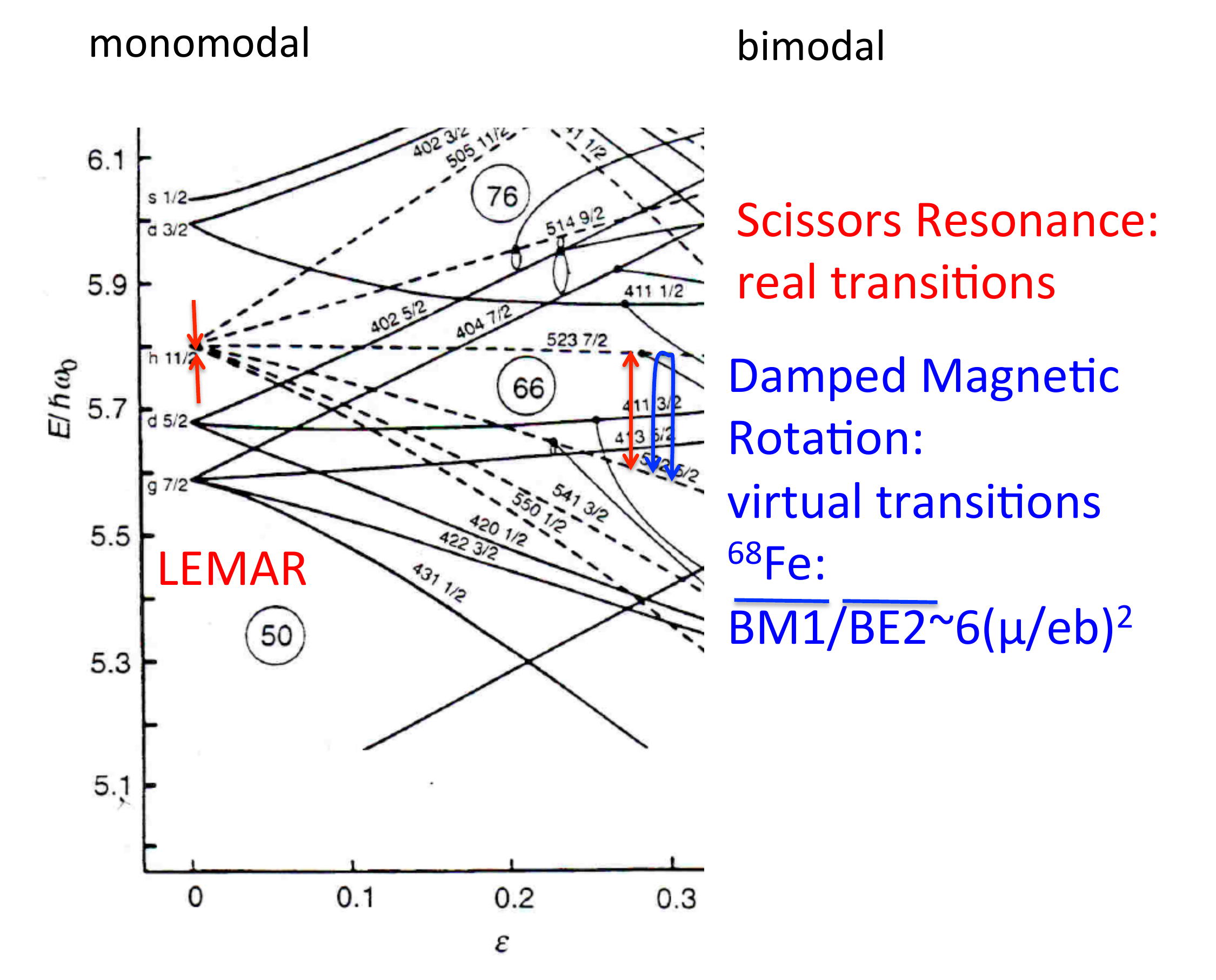}
\end{center}
 \caption{\label{f:Nilsson} (Color online) Transitions that generate the LEMAR spike and the Scissors Resonance. The energy
 scale is $\hbar \omega_0$=7.7 MeV for $A=$153.}
 \end{figure}

Fig. \ref{f:Nilsson} illustrates how the development of the bimodal LEMAR-SR structure can be attributed to the 
onset of stable quadrupole deformation above the yrast line when entering the open shell.
 The mechanism causing large
$B(M1)$ values  is reorientation of the high-j single-particle angular momenta. Without deformation this 
occurs  between the various configurations of the nucleons in an incompletely filled j-shell, $h_{11/2}$ in the example, 
which generates the  LEMAR spike as suggested in Ref. \cite{sch13} and discussed above. 
The magnetic (m) substates of the high-j multiplet split with the onset of deformation.  Reorientation 
occurs in two different ways. First, there are the real transitions between  the m-substates. Particle-hole
excitations of this type are known to generate the SR \cite{ham84}. As seen in Fig.  \ref{f:Nilsson},
the splitting between the m-states corresponds to the position of the SR for $\varepsilon=0.3$.  
Second, there are the virtual excitations from the occupied to the empty m-states. These are known
generating the rotational motion of deformed nuclei. Along 
a rotational band the angular momentum increases by a coherent gradual alignment of the spins of the nucleons  in the deformed orbitals, which is described by the 
virtual particle-hole excitations between the m-states. As discussed in our paper \cite{sch17}, the rotational coherence is only 
partially realized for  the transition between the states several MeV above yrast. With increasing deformation, the
LEMAR spike changes from incoherent thermal-like radiation to partially coherent rotational radiation.   The onset of 
damped rotation can be seen in the E2 strength function for $^{68}$Fe shown in Fig. \ref{f:f1Fe}.  The bump at 1.2 MeV is generated by 
stretched E2 transitions with $E_\gamma=2I/{\cal J}$, where ${\cal J} \approx 12$ MeV$^{-1}$ is close to the rigid body moment of inertia. 
The LEMAR spike of $^{68}$Fe in Figs. \ref{f:BM1Fe} and \ref{f:f1Fe} deviates from the exponential form at the lowest energies, which can be traced back
to the presence of an rotational component with  $E_\gamma=I/{\cal J}$. 

In $(\gamma,\gamma')$ experiments $1^+$ states are excited from the ground
state that group to the SR in deformed nuclei. In the case of $^{68}$Fe the summed  strength of these transitions is calculated to be    
 $\sum B(M1,0^+_1\rightarrow 1^+)= 1.7~\mu^2$.    The integral of the strength function between 2 and 5 MeV 
 amounts to 6.6 $\mu^2$, which is  three and a half times larger (see Fig. \ref{f:f1Fe}). From their experiment on $^{151,153}$Sm, Ref. \cite{sim16} assigned an integrated  strength
 of 7.8 $\mu^2$ to the SR, which is two and a half times larger that the summed strength of the excitation 
 of the SR from the ground state of $^{152}$Sm (see Fig. \ref{f:Sm153}). A similar enhancement of the SR strength in the $\gamma$ decay as 
 compared with the excitation of the SR from the ground state by $(\gamma,\gamma')$ experiments has 
 been reported before (see Ref. \cite{sim16} for references). 
 
 The origin of the enhancement of the transition strength
 is an increase of transitions from excited states as compared to $0^+_1\rightarrow 1^+$.
 For example in $^{68}$Fe, $\sum B(M1,0^+_2\rightarrow 1^+)= 3.4\mu^2$.  We attribute the larger summed strengths between excited states to a
suppression of the pair correlations with increasing excitation energy, i.e.
the thermal quenching of pairing. Pair correlations tend to couple the 
high-$j$ orbits to zero spin, which obstructs the reorientation that generates
the M1 radiation.

 S. F. acknowledges support by the DOE Grant DE-FG02-95ER4093, B. A. B.
support by the NSF Grant PHY-1404442, and all authors acknowledge the
importance of the ECT* workshop on "Statistical properties of nuclei",
July 11 - 15, 2016, for advancing this research.

\end{document}